\documentclass[twocolumn,showpacs,preprintnumbers,amsmath,amssymb]{revtex4}

\usepackage{epsfig}
\usepackage{amssymb}
\def \ra {\rightarrow}

\def \bc {\begin{center}}
\def \ec {\end{center}}
\def \be {\begin{equation}}
\def \ee {\end{equation}}
\def \bdm {\begin{displaymath}}
\def \edm {\end{displaymath}}
\def \bea {\begin{eqnarray}}
\def \eea {\end{eqnarray}}
\def \bac {\begin{array}{c}}
\def \eac {\end{array}}

\def \bmp {\begin{minipage}[h]{1\textwidth}}

\def \emp {\end{minipage}}

\def\figh{8cm} 
\def\sfigh{4.2cm} 

\def \dd {\mbox{d\raisebox{0.75ex}{\hspace*{-0.32em}-}\hspace*{-0.02em}}}

\def\Journal#1#2#3#4{{#1} {\bf #2} (#4) #3}

\def\NPA{{\em Nucl. Phys.} A}
\def\PLB{{\em Phys. Lett.} B}

\def\PRC{{\em Phys. Rev.} C}
\def\PRD{{\em Phys. Rev.} D}

\def\PS{\em Physica Scripta}

\begin{document}

\title{The detailed mechanism of the $\eta$ production in pp
           scattering up to the T$_{lab}\,$ = 4.5 GeV}

\author{Sa\v sa Ceci}\email{sasa.ceci@irb.hr}
 \author{Alfred \v Svarc}
 \author{Branimir Zauner}
\affiliation{%
\\Ru\protect\dd er Bo\v{s}kovi\'{c} Institute,
\\Bijeni\v cka Cesta 54, 10000
 Zagreb, Croatia}%


\date{\today}

\begin{abstract}
Contrary to very early beliefs, the experimental cross section
data for the $\eta$ production in proton-proton scattering are
well described if  $\pi$ and only $\eta$ meson exchange diagrams
are used to calculate the Born term. The inclusion of initial and
final state interactions is done in the factorization
approximation by using the inverse square of the Jost function.
The two body Jost functions are obtained from the $S$ matrices in
the low energy effective range approximation.  The danger of
double counting in the $p \eta$ final state interaction is
discussed. It is shown that higher partial waves in meson-nucleon
amplitudes do not contribute significantly bellow excess energy of
$Q=100\,$ MeV. Known difficulties of reducing the multi resonance
model to a single resonance one are illustrated.
\end{abstract}
\maketitle

\section{Introduction}

For a few decades the only way of detecting  the $\eta$ meson
signal in the intermediate energy proton-proton scattering was to
perform the pionic three-prong experiments  and finding the
characteristic invariant mass. In the CERN/HERA report
\cite{Fla84} a compilation of all mutually normalized and
otherwise adjusted experimental data for the $pp\ra pp \eta$
process from T$_{\mathrm{lab}}\,$ = 2 GeV to T$_{\mathrm{lab}}\,$
= 13 GeV have been presented. Yet, first theoretical models
\cite{Ger90,Lag91,Vet91} appeared at the beginning of the 90s,
accompanied by the first published experimental results
\cite{Ber93}. The common denominator of these first models is a
similar reaction mechanism: the $\eta$ meson in two proton
collisions is produced when an intermediate meson, emitted in the
meson-production  vertex, interacts with a proton forming an $N^*$
resonance which decays into $\eta$ meson and proton. Everything
else varies from model to model: the number and type of
intermediate mesons ($\pi$, $\eta$, $\rho$, $\omega$, $\sigma$,
...), the type of excited state (a resonance formation or some
re-scattering in addition), etc. The contribution of $\rho$ meson
exchange is recognized as a dominant one, or at least of equal
importance as the $\pi$ meson exchange. The influence of $\eta$
exchange has been considered as insignificant. The higher order
terms expressed through different forms of initial and final state
interactions had to be included, and it has been shown that they
play a crucial role for the reliability of the calculation. The
agreement of these models with the experimental data has been
obtained by adjusting the free coupling constants. The new
measurements have soon appeared \cite{Chi94,Cal96,Cal98,Cal99},
and the old models have been improved accordingly
\cite{Wil93,Fae96,San98,Ged98}. The number and type of exchanged
mesons still varied. \\ \\A model  has been proposed \cite{Bat97}
where the meson-nucleon partial-wave T-matrices obtained in the
multi-resonance, coupled channel and unitary model \cite{Bat9598}
are used in the $\eta$-meson production vertex instead of
individual $N^*$ resonances. The discussion which intermediate
mesons dominate the process is reduced to calculating the
pp$\rightarrow$ pp$\eta$ process with the predetermined relative
strength of the two-body $eta$-meson production T-matrices for
various opened channels. Contrary to former statements, it has
been shown that a good agreement with the experimental results for
the total cross section is achieved using only $\pi$ and $\eta$
contributions, without any need to invoke heavier mesons.\\
\\New low energy experiments reporting angular correlation
measurements have been done at J\" ulich \cite{Smy99,Mos03},  so
theoretical focus has moved from evaluating total cross section to
more detailed calculations \cite{Nak03}.\\ \\This article offers
an improvement of the previous model \cite{Bat97} and the
extension of the examined energy range. Basic conclusions of the
first model are confirmed: the $\pi$ and $\eta$ meson exchange
contributions are sufficient for the Born term so there is no
demonstrable necessity to introduce other meson contributions; the
initial and final state interactions are essential for the shape
and size of the cross section. The new, formerly unreachable
conclusions presented in this article are enabled by a more
transparent treatment of the initial and final state interactions.
This model reasonably well describes the total cross section for a
wide energy range from very near threshold to {
T$_{\mathrm{lab}}\,$ = 4.5 GeV}.

\maketitle
\section{Formalism}

\begin{figure}[!h]\bc
\epsfig{figure=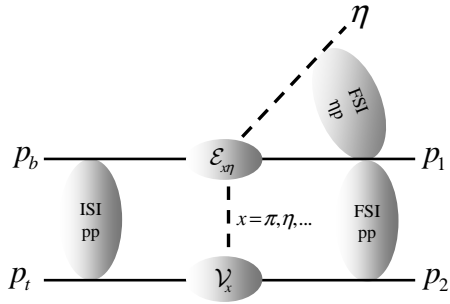,height=4cm}
\caption{ Ingredients of the  model. ISI and FSI represent the
initial and final state interactions. $\mathcal{V}_x$ is the Bonn
vertex, while the $\mathcal{E}_{x\eta}$ is the $\eta$ production
amplitude.}\label{fig:1}
\ec\end{figure}

The ingredients of the model are presented in Fig. \ref{fig:1}.
The differential cross section of the full process is given as:
\be \label{LipsXs} d\sigma=\frac{1}{\mathcal{F}}f_{out} \,
\overline{\left|\sum_{x}\mathcal{E}_{x\eta}\,\mathcal{P}_x
\,\mathcal{V}_x \right|^2} f_{in} \, d\Phi\,\ee where
$\mathcal{F}$ is a standard flux factor and $d\Phi$ is the
three-body Lorentz invariant phase space. The three-particle
vertex $\mathcal{V}_x$ is completely determined by the Bonn
potential parameters \cite{Machleidt}, modified by keeping the
momenta in the propagator and form factors relativistic.
Pseudoscalar meson-proton coupling constants and cut-off
parameters, used in this calculation, are identified with the
corresponding on mass shell meson-nucleon values from the same
reference: $g_{\pi NN}^2/4\pi=13.6$, $g_{\eta NN}^2/4\pi=3$,
$\Lambda_\pi=1.3$ GeV and $\Lambda_\eta=1.5$ GeV.

Eta emission vertex $\mathcal{E}_{x\eta}$ is given by $xp\ra \eta
p$ scattering amplitudes from our partial wave analysis
\cite{Bat9598}. Off-shell amplitudes have been constructed by
keeping the multi resonant $T$-matrices on shell and extrapolating
partial wave projectors, following the recipe given in
\cite{Kroll}.

The initial and final state correction factors have roots in a
threshold transition from the initial two proton $^{33}P_1$ state
to the final two proton $^{31}S_0$ state with the $\eta p$
subsystem in the s-state.

The final state correction factor is given in the factorization
approximation using the two body Jost functions in the low energy
effective range approximation \cite{Goldberger}:
\be \label{ISI,FSI} 
 f_{out}=\frac{1}{|\mathcal{J}(p_{12})|^2}
\frac{1}{|\mathcal{J}(p_{1\eta}) \mathcal{J}(p_{2\eta})|
^{2n}},\,\, \ee where two body relative momentum is defined as
$p_{AB}=|{\bf p}_A-{\bf p}_B|$/2 with $A$ and $B$ as pairs of the
final state particles. The coefficient $n$ is a part of our model.
Allowing the possibility that the scattering length $a$ and
effective range $r$ are complex numbers, the two body correction
factor used for the evaluation of final state contribution - Eq.
(\ref{ISI,FSI}) is, in the s-wave low energy effective range
approximation, generically given as: \be \label{FSIrelation}
\frac{1}{|\mathcal{J}(p)|^2}=\frac{p^2-2\,p\,\mathrm{Im}\,
\alpha_+ +|\alpha_+|^2}{p^2+2\,p\,\mathrm{Im}\,\alpha_-
+|\alpha_-|^2} \ , \ee where $\alpha_ \pm = \frac{1}{r} \pm
\sqrt{\frac{1}{r^2} +\frac{2}{a\,r}}$.

To calculate the Jost functions we have used the following $pp$
scattering length and effective range parameters for the two
protons in the $^{31}S_0$ state: $ a^{  pp}_{0}=\,7.8$ fm and $r^{
pp}_{0}=\, 2.79$ fm \cite{Sto99}.

For the $\eta p$ $S_{11}$ subsystem, the s-wave scattering length
is given by $a^{p\eta}_{0}=(0.717+i\, 0.265)$ fm \cite{Bat9598}.
For this article, the corresponding effective range
$r^{p\eta}_{0}=(-1.574-i\, 0.020)$ fm has been obtained by fitting
the $S_{11}$ wave $T$-matrix from the same reference.

For the $^{33}P_1$ state, the Jost function has been estimated by
using p-wave effective range relation \cite{Goldberger}. The used
values are $a^{\ pp}_{10}=-3.3$ fm$^3$ and $r^{ \ pp}_{10}=\,4.22$
fm$^{-1}$ \cite{Bergervoet}.

The initial state factor $f_{in}$ is indistinguishable from unity
for energies above $\eta$ production threshold, while final state
correction factor exhibits the distinct variations in magnitude.

\section{Assembling the model}

We have calculated  the Born term for this process using the
$S_{11}$ meson-nucleon amplitudes of ref. \cite{Bat9598} for all
three channels: $\pi$, $\eta$, and the "effective" two body
channel. It turns out that the individual contributions of $\pi$
and $\eta$ mesons are comparable in size and slightly dominated by
the first one. The "effective" meson contribution is negligible
\cite{Bat97,Cec02}. Therefore, from now on the third meson
contribution (in which the $\rho$ contribution is implicitly
included) will be disregarded in this work.

To determine the relative sign of the remaining two meson
contributions, we have to obtain the qualitative and quantitative
agreement of the model predicted total cross section with all
experimental data available within the low and intermediate energy
range. We do it in three steps: we calculate the Born term
contribution (model A), add the $pp$ final state interaction
(model B), and finally include the $p\eta$ final state
interaction. We offer two versions of the $p\eta$ final state
interaction: the full Jost function ($n=1$ in Eq. (\ref{ISI,FSI})
- model C), and an arbitrary effective reduction of the final
state interaction because of possible double counting in the
$\eta$ meson production vertex when the full $T$-matrices are used
instead of individual N$^*$ resonances ($n=\frac{1}{2}$ in Eq.
(\ref{ISI,FSI}) - model D.)

The agreement of the model predictions with the compilation of the
experimental data points to the appropriate model.

\section{Results and conclusions}
\subsection{Total cross section}

\begin{figure}[!h]\bc
\epsfig{figure=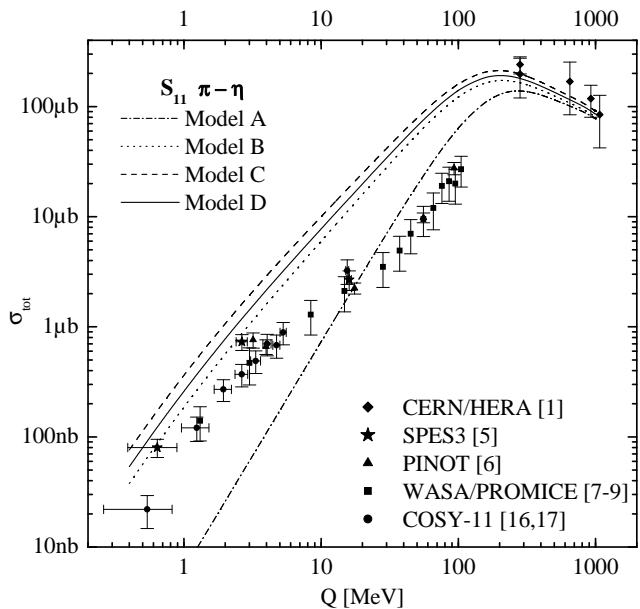,height=\figh,angle=0}
\caption[Fig.2.]{The total $pp\ra pp\eta$  cross section. {
$S_{11}$ partial wave in multi resonant model is only used. The
negative $\pi$ vs. $\eta$ relative sign is chosen. The
interference is  constructive. Models A,B,C and D are described in
the text. $Q$ is the excess energy.}}\label{fig:2}
\ec\end{figure}

\begin{figure}[!h]\bc
\epsfig{figure=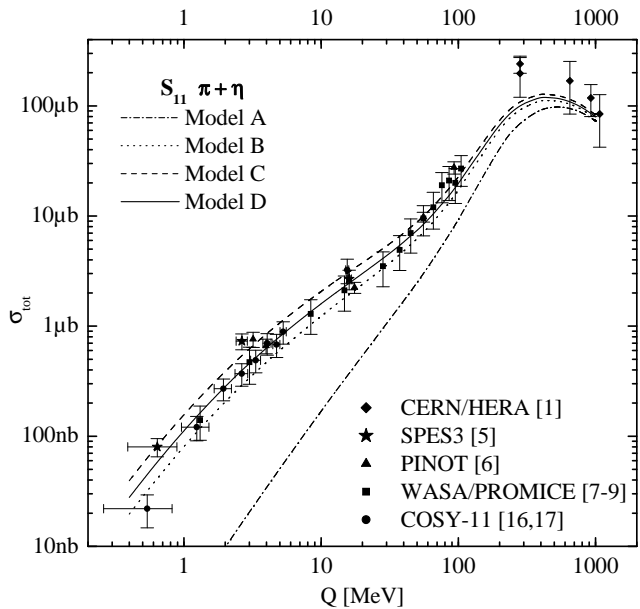,height=\figh,angle=0}
\caption[Fig.3.]{The total $pp\ra pp\eta$  cross section. {
$S_{11}$ partial wave in multi resonant model is only used. The
positive $\pi$ vs. $\eta$ relative sign is chosen. The
interference is destructive. Models A,B,C and D are described in
the text. $Q$ is the excess energy.}}\label{fig:3}
\ec\end{figure}

The comparison of model predictions with the compilation of
experimental cross section data for different $\pi$ vs. $\eta$
relative signs is given in Fig. \ref{fig:2}  and Fig. \ref{fig:3}.
 The negative $\pi$ vs. $\eta$ relative sign, as
shown in Fig.\ref{fig:2}, is ruled out.

The inclusion of the  $pp$ final state interaction improves the
agreement of the model with the experimental data significantly,
as can be seen in Fig. \ref{fig:3}.
Used model for the $p\eta$ FSI gives the overall agreement between
the model and experiment (see dashed and full lines in Fig.
\ref{fig:3}). Based on the results presented in Fig. \ref{fig:3}
we are in favor of model D. However, as the precision of the
measured data in the excess energy range of 10 MeV $<Q<$ 100 MeV
is not adequate (normalization problems), we can only suspect that
the double counting effects are hidden in the proposed formalism.

\begin{figure}[!h]\bc
\epsfig{figure=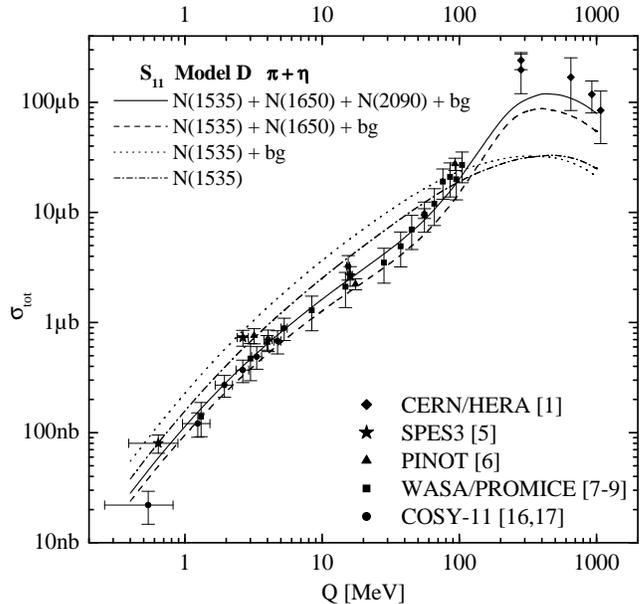,height=\figh,angle=0}
\caption[Fig4]{The total $pp\ra pp\eta$ cross section when
individual $S_{11}$ wave resonances are consecutively added to
form the two body $T$-matrix. { The resonance nomenclature is
taken over from PDG.}}\label{fig:4}
  \ec\end{figure}

The consecutive addition of individual $S_{11}$ wave resonances in
forming the partial wave $T$-matrix is shown in Fig. \ref{fig:4}.
Results presented in this figure  confirm our repeatedly made
statement that a single resonance model drastically fails in
calculating the total cross section. We find out that {\it only }
the inclusion of {\it all three} resonances reproduces the
measured cross section in the considered energy range.

\begin{figure}[!h] \bc
\epsfig{figure=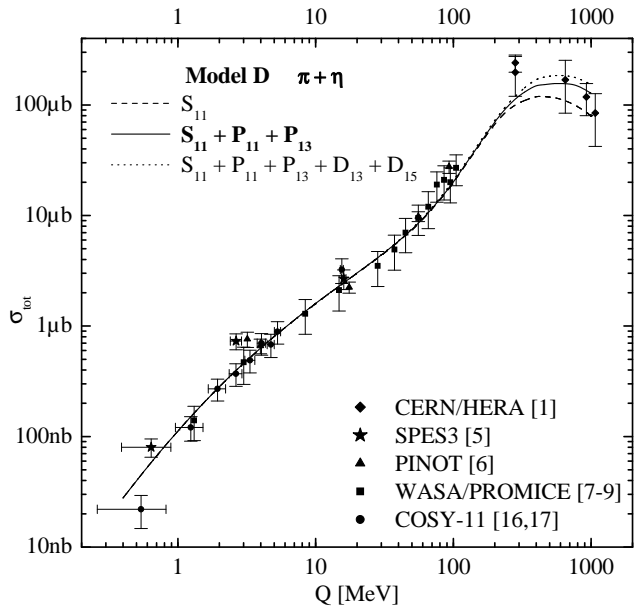,height=\figh,angle=0}
\caption[Fig5]{The total  $pp\ra pp\eta$ cross section. { In
addition to the  S-wave, P and D waves  are as well used in $xN\ra
\eta N$ $T$-matrices}}\label{fig:5}
 \ec\end{figure}

In Fig. \ref{fig:5} the higher partial-waves are included into the
two-body multi resonant $T$-matrices with the intention of
obtaining better agreement with experiment at energies higher then
Q = 100 MeV. The inclusion of higher partial-waves into the $\pi
N$ $T$-matrices start to be noticeable only at surprisingly high
energies (above Q $\approx$ 300 MeV). That indicates that the
improvement of the present model should be directed to a more
rigorous treatment of the initial and final state interaction
($pp$ in particular) prior to the possible improvement of the Born
term.

Additional experiments in the intermediate energy region (100 MeV
$<$ Q $<$ 1 GeV) would be greatly appreciated. Proper ISI and FSI
treatment, along with extending the experimental data set,
might increase the reliability of the resonant parameters,
the $S_{11}$ in particular.

\subsection{Differential cross sections}
We have already determined  the ingredients of the model in such a
way  that the shape and size of the experimental total cross
section values are fairly well reproduced. The angular
distributions therefore come out as a prediction. Comparing the
predictions of model D with the known experimental values of
differential cross sections represent an additional test of
understanding the physical nature of the process.

\begin{figure}[ht!]
\epsfig{figure=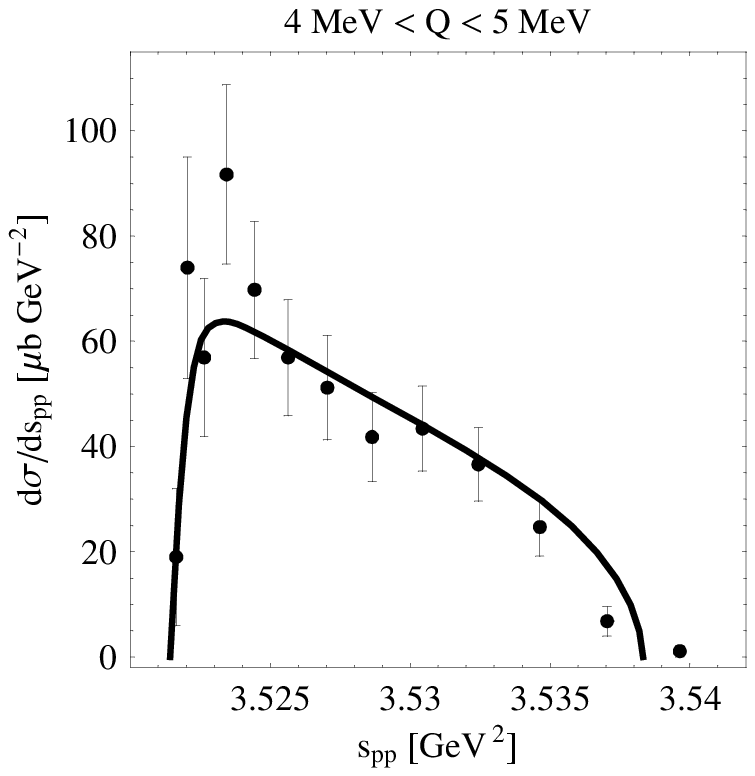,height=\sfigh,angle=0}\\
\epsfig{figure=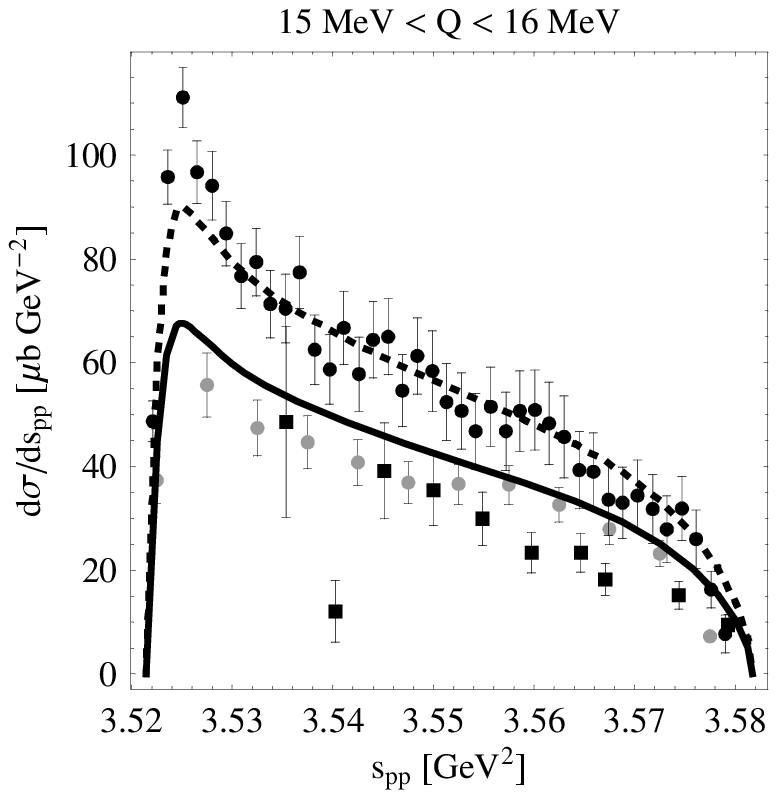,height=\sfigh,angle=0}
\epsfig{figure=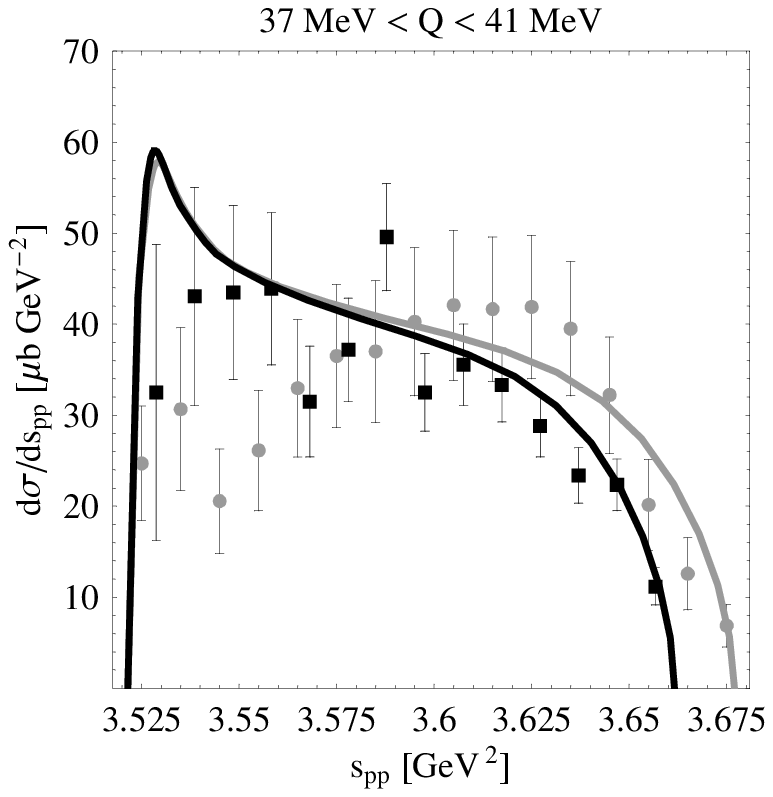,height=\sfigh,angle=0}\\
\epsfig{figure=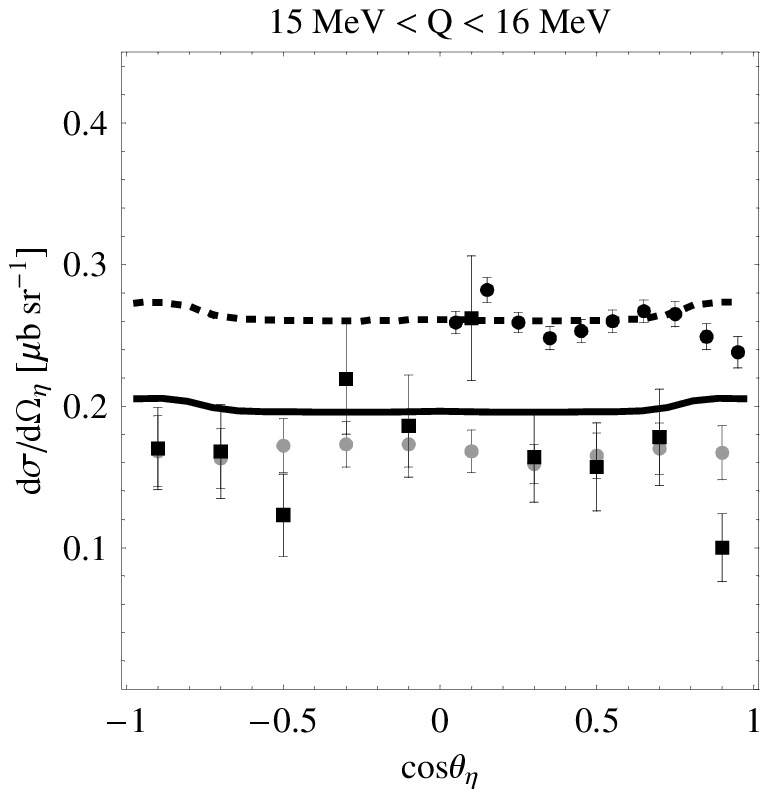,height=\sfigh,angle=0}
\epsfig{figure=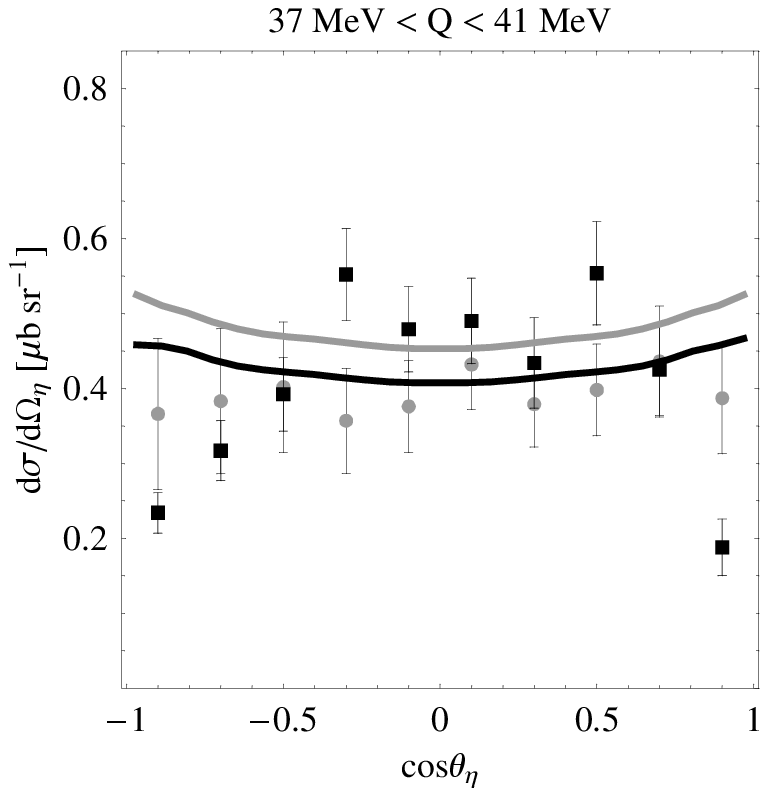,height=\sfigh,angle=0}\\
 \caption{The
differential cross sections. Data are from Cal\' en
\protect\cite{Cal99} (black squares), Abdul-Bary
\protect\cite{Abd02} (gray circles), and Moskal
\protect\cite{Mos03} (black circles). The black full lines are
predictions of model D of this article for excess energies 4.5
MeV, 16 MeV, and 37 MeV respectively. The black dashed line is
model D prediction for 16 MeV normalized by the factor 1.35. The
gray line is model D prediction for 41 MeV.}\label{fig:6}
\end{figure}

Since the final state consists of three particles, one can
construct and measure several partially integrated differential
cross sections. The $pp\eta$ system in the final state is
described by 12 variables and 7 of them are fixed by
energy-momentum conservation and particle masses. The differential
cross sections with respect to the 5 remaining quantities can be
measured. We have compared the predictions of our model with the
experimental values for the  angular distribution
$d\sigma/d\Omega_\eta$, where $\Theta _\eta$ is the angle of the
$\eta$ momentum in the center-of-mass frame with respect to the
beam axis,  and the invariant mass distribution $d\sigma/ds_{pp}$
where $s_{pp}$ is the proton-proton invariant mass. Measurements
of these quantities have been recently reported by Calen
\cite{Cal99} at CELSIUS facility for  $Q = 15 \ {\rm and} \ 37$
MeV , Moskal \cite{Mos03} at COSY-11 for $Q = 15.5$ MeV , and
Abdul-Bary \cite{Abd02} $Q = 16 \ {\rm and} \ 41$ MeV at COSY TOF
(time of flight) detector. The Abdul-Bary \cite{Abd02} experiment
as well gives the value of $d\sigma/ds_{pp}$  at the $Q = 4.5$
MeV.  The differential cross sections with respect to the
remaining two variables of interest (proton-$\eta$ invariant mass
and the angle of relative momentum of the proton-proton system to
the beam axis) have been measured, but are not evaluated in this
article. The fifth variable, the azimuthal orientation, is of no
significance in the unpolarized measurements.

Before we start the comparison we feel compelled  to comment the
overall agreement of existing experiments. Out of three
experiments, two of them (Cal\' en \cite{Cal99} and Moskal
\cite{Mos03})  report quantitative measurements, while the third
one (Abdul-Bary \cite{Abd02}) gives results  in arbitrary units
only, and normalizes to Cal\' en \cite{Cal99}. The two
quantitative experiments are clearly in strong disagreement at
nearby energies: Cal\' en at $Q=37$~MeV \cite{Cal99} vs.
Abdul-Bary at $Q=41$~MeV  \cite{Abd02}. It is interesting to
mention that initially reported values of the COSY-11 experiment
\cite{Mos02} are very similar to the Cal\' en \cite{Cal99}.
However, in the currently valid version of the measured data the
calibration have been increased by $\approx$ 40 \% at $Q=
15.5$~MeV, but remained within the error bar for the $Q = 4.5$ MeV
\cite{Mos03}. So, we are left with the situation that we can just
compare the shape of differential cross sections paying less
importance to the magnitude. Disregarding the absolute
normalization we see the fair agreement of the shape of the
differential cross section for all experiments at $Q  \approx 15$
MeV. At  $Q \approx 41$ MeV some discrepancies remain: the Cal\'
en \cite{Cal99} data show the distinct structure in
$d\sigma/d\Omega_\eta$ which is missing in ref. \cite{Abd02}, and
the shape of $d\sigma/ds_{pp}$ is notably different in both
experiments. However, the differences remain within error bars. It
is therefore clear that the detailed quantitative comparison of
theory and experiment can not be done and that achieving the
quantitative agreement awaits for the experimental accordance.

In Fig. \ref{fig:6}  we show the  results of our calculations
compared to the world data collection. The full lines give the
comparison of model D predictions with experiment. The present
model nicely reproduces the strong dynamically created departure
of both differential cross sections from the symmetric, phase
space induced shape. The agreement in shape and size is good for
all energies, with the exception of $Q=15.5$ MeV value of Moskal
(COSY-11) \protect\cite{Mos03}. However, if we normalize our model
predictions by 35 \% (Fig. \ref{fig:6} - dashed line) the data are
again reproduced very nicely. For the $d\sigma/ds_{pp}$ the low
$s_{pp}$ peaking is nicely followed for all three energies, but
the more pronounced structure at $s_{pp}\approx 3.63$~GeV$^2$ in
the Abdul-Bary data \cite{Abd02} for $Q=41$~MeV  is not
understood. The same agreement in general trend is achieved for
the $d\sigma/d\Omega_\eta$ but the Cal\'  en structure at
$Q=37$~MeV is not accounted for.

\section{Summary}
Within the framework of the proposed model we show that the single
resonance model, using only N(1535), drastically fails to describe
the experimental data. Next $S_{11}$ resonance, N(1650), has to be
included in any model to obtain the shape of the total cross
section. The inclusion of the third, controversial, N(2090)
$S_{11}$ resonance represents a further improvement. The inclusion
of P-waves is essential for the energies \mbox{100 MeV $<$ Q $<$ 1
GeV}. All these contributions are effectively incorporated in the
proposed model. The FSI corrections are essential to obtain the
qualitative and quantitative agreement with experiment. The
present way how the FSI is treated in this article, namely using
only the S-waves  for the Jost  function in the low energy
effective range approximation, is enough to understand the data to
the present level of experimental precision. The improvement of
the data, the differential cross sections in particular, will
require the improvement of the FSI treatment as well.

\end{document}